\documentclass[rnote]{aa}

\usepackage[utf8]{inputenc} 
\usepackage[T1]{fontenc}

\usepackage{color} 
\usepackage[breaklinks=true]{hyperref} 

\usepackage{aas_macros}
\usepackage{natbib,twoopt}
\bibpunct{(}{)}{;}{a}{}{,} 

\usepackage[toc,page,titletoc, title]{appendix}

\usepackage{pdfpages}
\usepackage{graphicx}
\usepackage{wrapfig}
\usepackage{subfig}
\usepackage{tabularx,array}

\newcolumntype{G}{!{\vrule width 1pt}}
\usepackage{hhline}
\usepackage{multirow}
\usepackage{slashbox}

\usepackage{arydshln}
\usepackage{amsmath}
\usepackage{amsfonts}
\usepackage{amssymb}
\usepackage{marvosym}

\usepackage{textcomp}

\usepackage{mathtools}
\usepackage{cases}
\usepackage{empheq}
\usepackage{slashbox}

\DeclareSymbolFont{boldsymbols}{OMS}{cmsy}{b}{n}
\DeclareSymbolFontAlphabet{\mathbfcal}{boldsymbols}

\newcommand{\mcal}{\mathcal}

\newcommand{\mbf}{\mathbf}
\newcommand{\mrm}{\mathrm}

\newcommand{\Imcro}[1]{\mathcal I\!m \left[ #1 \right]}

\newcommand{\ie}{\leqslant}

\newcommand{\intervalleR}[3]
 { #1 \in \textrm{[}#2,#3\textrm{]} }
 
\begin{document}

\title{The surface signature of the tidal dissipation of the core in a two-layer planet}
\titlerunning{Surface signature of the tidal dissipation of the core}

\author{F. Remus\inst{1, 2, 3}, S. Mathis\inst{3, 4}, J.-P. Zahn\inst{2}, V. Lainey\inst{1}}
\offprints{F. Remus}

\institute{
 IMCCE,  Observatoire de Paris -- UMR 8028 du CNRS -- Universit\'e Pierre et Marie Curie, 77 avenue Denfert-Rochereau, F-75014 Paris, France
 \and
 LUTH, Observatoire de Paris -- CNRS -- Universit\'e Paris Diderot, 5 place Jules Janssen, F-92195 Meudon Cedex, France
 \and
 Laboratoire AIM Paris-Saclay, CEA/DSM -- CNRS -- Universit\'e Paris Diderot, IRFU/SAp Centre de Saclay, F-91191 Gif-sur-Yvette, France
 \and
 LESIA, Observatoire de Paris -- CNRS -- Universit\'e Paris Diderot -- Universit\'e Pierre et Marie Curie, 5 place Jules Janssen, F-92195 Meudon, France
 \\{}\\{}
 \email{francoise.remus@obspm.fr, stephane.mathis@cea.fr, jean-paul.zahn@obspm.fr, lainey@imcce.fr} 
}

\date{Received ; accepted}

 
  \abstract 
   {	
    Tidal dissipation, which is directly linked to internal structure, is one of the key physical mechanisms that drive systems evolution and govern their architecture. 
   A robust evaluation of its amplitude is thus needed to predict evolution time for spins and orbits and their final states.
   }
   {	
   The purpose of this paper is to refine recent model of the anelastic tidal dissipation in the central dense region of giant planets, commonly assumed to retain a large amount of heavy elements, which constitute an important source of dissipation.
   }
   {	
   The previous paper evaluated the impact of the presence of the static fluid envelope on the tidal deformation of the core and on the associated anelastic tidal dissipation, through the tidal quality factor $Q_c$.
   We examine here its impact on the corresponding effective anelastic tidal dissipation, through the effective tidal quality factor $Q_p$.
   }
   {	
   We show that the strength of this mechanism mainly depends on mass concentration.
   In the case of Jupiter- and Saturn-like planets, it can increase their effective tidal dissipation by, around, a factor $2.4$ and $2$ respectively.
   In particular, the range of the rheologies compatible with the observations is enlarged compared to the results issued from previous formulations.
   } 
   {
   We derive here an improved expression of the tidal effective factor $Q_p$ in terms of the tidal dissipation factor of the core $Q_c$, without assuming the commonly used assumptions.
   When applied to giant planets, the formulation obtained here allows a better match between the anelastic core's tidal dissipation of a two-layer model and the observations.
   }



\maketitle

\section{Introduction and motivations}
\label{Sect:intro}

\defcitealias{RMZL..2012}{RMZL12}
In \cite{RMZL..2012} (hereafter \citetalias{RMZL..2012}), we have studied the tidal dissipation of the anelastic core of a two-layer planet.
In that model, the core was assumed to be the main source of tidal dissipation.
The envelope was considered as a non-viscous fluid, which was sensitive to the tidal deformation only through its action on shape deformation.
The paper derived the tidal dissipation of the core, quantified by the ratio $k_2^c/Q_c$, where $k_2^c$ is the norm of the complex Love number of second-order at the mean surface of the core ($\tilde{k}_2(R_c)$) and $Q_c$ is its tidal quality factor.
Nonetheless, the equations that describe the dynamical evolution of a given satellite are not directly expressed with $\Imcro{\tilde{k}_2(R_c)}=k_2^c/Q_c$.
According to \cite{Kaula..1962}, for example, the evolution rate of the semi-major axis ${\mrm da}/{\mrm dt}$ is proportional to $R_p^5 \, {k_2^p}/{Q_p}$,
where we use here "$p$" sub- and super-scripts which relate to the surface of the planet.
Thus, the quantity that intervenes here is the global tidal dissipation, also called the effective tidal dissipation, {\it i.e.}
the imaginary part of the complex Love number taken at the surface of the planet.\newline
As mentioned in \citetalias{RMZL..2012}, recent studies managed to extract the tidal dissipation of Jupiter \citep{Lainey.etal..2009} and Saturn \citep{Lainey.etal..2012} from astrometric data.
This quantity corresponds to the ratio ${k_2^p}/{Q_p}$, assuming that $R_p$ and the other parameters are known.
Then, the question is: which is the tidal dissipation in the core able to account for the observed dissipation?
In other words, if one considers, as done in \citetalias{RMZL..2012}, that the source of tidal dissipation is restricted to the core, it seems obvious that one has to relate the tidal dissipation in the core to the effective tidal dissipation of the planet which can be directly compared to the observations.
The purpose of this paper is to explicitly derive such relation.
We will first discuss in sec.~\ref{Sect:Qeff} the limits of application of the standard formulation, and its implicit consequences.
Then, we will show how to derive the effective tidal dissipation, knowing the tidal dissipation of the core, and taking into account both the overload and the gravitational attraction exerted by the envelope on the core.
Finally, sec.~\ref{Sect:Jup_Sat} evaluates the strength of this mechanism in gas giant planets.

\section{Effective tidal dissipation of a two-layer planet}
\label{Sect:Qeff}

As assumed in \citetalias{RMZL..2012}, we will consider a two-layer planet of radius $R_p$, having an homogeneous density $\rho_c$ up to the radius $R_c$ at which a density jump occurs down to $\rho_o$, then staying constant up to the surface of the planet. 

\subsection{The standard approach}
\label{SubSect:Qeff_standard}

Except \cite{Dermott..1979} and recently \citetalias{RMZL..2012}, none of the studies dealing with the anelastic tidal dissipation of planets or satellites had taken into account the role of an outer layer surrounding the anelastic region as a non-negligible factor that could enhance the tidal dissipation of the core.
Thus, so far, since the envelope had no influence on the core, it could be omitted. The problem could then be reduced to seeking the tidal dissipation of an homogeneous planet of radius $R_c$ and density $\rho_c$ which would induce the same satellite migration as the supposed two-layer planet of radius $R_p$.
In that case, such a quantity would obey, as commonly assumed:
\begin{equation}
\label{Eq:k2Qc_STD}
	\frac{k_2^p}{Q_p} = \left( \frac{R_c}{R_p} \right)^5 \frac{k_2^c}{Q_c} \;.
\end{equation}
Nevertheless, such formulation no longer holds if we consider both the gravitational attraction and the overload exerted by the tidally deformed envelope on the core.

\subsection{Retroaction of the static envelope}
\label{SubSect:Qeff_bicouche}

In this section, we will show that considering these effects, as done in \citetalias{RMZL..2012}, Eq. \ref{Eq:k2Qc_STD} is no longer valid.
According to Eq. 6 of \cite{Dermott.etal..1988},
\begin{equation}
\label{Eq:k2Qc_bicouche}
	\frac{k_2^p}{Q_p} = \left( \frac{R_c}{R_p} \right)^5 F_p^2 \frac{k_2^c}{Q_c} \;.
\end{equation}
In this paper, the authors referred to \cite{Dermott..1979} for the derivation of the factor $F_p$, which {\it "is a factor that allows for the enhancement of the tide in the core by the tide in the overlying ocean and for the effects of the density contrast between the core and the ocean"}.

We will derive here an explicit expression of the factor
\begin{equation}
\label{Eq:Gp_def}
	G_p \equiv F_p^2 = \left( \frac{R_p}{R_c} \right)^5 \frac{\Imcro{\tilde{k}_2 (R_p)}}{\Imcro{\tilde{k}_2 (R_c)}} \;,
\end{equation}
thanks to \citetalias{RMZL..2012}, where we discussed the weaknesses of Dermott's formulation for the tidal dissipation of the core appearing in the paper of \citeyear{Dermott..1979}.

The complex Love number is given by $\tilde{k}_2 (R_\Lambda) = \tilde{\Phi}'(R_\Lambda)/U(R_\Lambda)$, where $U$ designs the tidal potential, $\tilde{\Phi}'$ the perturbed gravitational potential, and the $\Lambda$ subscripts stand either for $c$ (for core surface quantities) or $p$ (for planet surface quantities).
The tidal potential is of the form
\begin{equation}
\label{U_def}
	U(\mbf r ) = - \zeta_c g_c \frac{r^2}{R_c^2} P_2(\cos\Theta) \;,
\end{equation}
where $g_c$ is the gravity induced by the core at its surface, and
\begin{equation}
	\zeta_c = \frac{m}{M} \left( \frac{R_c}{a} \right)^3 R_c
\end{equation}
is the tidal height at the surface of the core, with $m$ the mass of the perturber, and $a$ the semi-major axis of its orbit.
Let us denote the core and planet surfaces by, respectively:
\begin{align}
\label{Eq:sc_sp_def}
	\mrm s_c &= R_c \left[ 1 + S_2 P_2 (\cos\Theta) \right] \;, \\
	\mrm s_p &= R_p \left[ 1 + T_2 P_2 (\cos\Theta) \right] \;,
\end{align}
where we have used the polar coordinates $(r, \Theta)$ to locate a point P, $r$ being the distance to the center of the planet and $\Theta$ the angle formed by the radial vector and the line of centers, and where $P_2$ is the Legendre polynomial of second order.
Since there is no physical reason for which the surfaces of the core and the envelope suffer the same deformation, in general, $T_2 \neq S_2$.
The way these quantities are linked depends on the effective forces acting on the surface of the core.
From here on, we will denote by $\varepsilon$ the ratio defined by
\begin{equation}
\label{Eq:beta_def}
	T_2 = \varepsilon S_2 \;.
\end{equation}
We first consider a purely elastic core.
The self-gravitational potential of the planet is the sum of the core and envelope contributions.
In the envelope (where $s_c \ie r \ie s_p$), they are  given by eqs. (49, 50) of \citetalias{RMZL..2012}:
\begin{subequations}
\label{Eq:Phic_Phip_def}
\begin{flalign}
	\Phi_c (\mbf r) &= - {g_c}{R_c} \left[ \frac{R_c}{r} + \frac{3}{5} \left( \frac{R_c}{r} \right)^3 S_2 P_2 \right] \;, \\
	\Phi_o (\mbf r) &= 
			\begin{multlined}[t] 
				- {g_c}{R_c} \frac{\rho_o}{\rho_c} \left[ 
							 \frac{3 R_p^2 - r^2}{2 R_c^2} + \frac{3}{5} \left( \frac{r}{R_c} \right)^2 T_2 P_2  \right.  \qquad \\
							 \left. - \frac{R_c}{r} - \frac{3}{5} \left( \frac{R_c}{r} \right)^3 S_2 P_2 \right]\;. 
			\end{multlined}
\end{flalign}
\end{subequations}
Thus, the effective deforming parts are:
\begin{subequations}
\label{Eq:Phi'c_Phi'p_def}
\begin{equation}
	\Phi'\left( R_\Lambda \right) = -g_c R_c \left( \frac{R_\Lambda}{R_c} \right)^2 Z'_\Lambda S_2 P_2(\cos\Theta) \;,
\end{equation}
where
\begin{equation}
	Z'_\Lambda = \frac{3}{5} \left[ \frac{\rho_o}{\rho_c} \varepsilon + \left( 1 - \frac{\rho_o}{\rho_c} \right) \left( \frac{R_\Lambda}{R_c} \right)^5 \right] \;.
\end{equation}
\end{subequations}
Therefore:
\begin{equation}
\label{Eq:k2_form_gen}
	k_2(\mrm R_\Lambda) \equiv \frac{\Phi(\mrm R_\Lambda)}{U(\mrm R_\Lambda)} = \frac{R_c}{\zeta_c} Z'_\Lambda S_2 \;.
\end{equation}
The condition that the surface of the planet $\mrm s_c$ is an equipotential of the total field $\Phi+U$ leads to
\begin{equation}
\label{Eq:zetaRc}
	\frac{\zeta_c}{R_c} = \frac{2}{5} \frac{\rho_o}{\rho_c} \left( \alpha \varepsilon - \beta \right) S_2 \;,
\end{equation}
where
\begin{equation}
\label{Eq:alpha_beta_def}
	\alpha = 1 + \frac{5}{2} \left( \frac{\rho_c}{\rho_o} - 1 \right)  \left( \frac{R_c}{R_p} \right)^3
	\textrm{; } 
	\beta = \frac{3}{5} \left( \frac{R_c}{R_p} \right)^2 (\alpha-1) \;.
\end{equation}
We finally obtain the general form of the Love numbers, generalized to the case of a two-layer planet:
\begin{equation}
\label{Eq:k2_epsilon_form}
	k_2(R_\Lambda) = \frac{5}{2} \frac{\rho_c}{\rho_o} \frac{Z'_\Lambda}{\alpha \varepsilon - \beta} \;.
\end{equation}

All the unknown quantities require the expression of $\varepsilon$ to be completely determined.
As mentioned before, it depends on the forces acting on the surface of the core.
According to \citetalias{RMZL..2012}, the radial displacement at the surface of the core $\xi_{r}(R_c) = R_c S_2 P_2 (\cos\Theta)$ is linked to the total normal traction $T_N(R_c) = X P_2 (\cos\Theta)$ applied on it, following
\begin{equation}
\label{Eq:radial_displacement}
	S_2 P_2 (\cos\Theta) = \frac{\xi_r(R_c)}{R_c} = \frac{5}{19 \mu_c} T_N(R_c) = \frac{5}{19 \mu_c} X P_2 (\cos\Theta) \;,
\end{equation}
where $\mu_c$ is the shear modulus of the core.
This traction corresponds to the normal stress acting on the surface of the core, and has to take into account both the gravitational forces and the solid and fluid loads.
Therefore $X$ is given by Eqs. (41) of \citetalias{RMZL..2012}
\begin{equation}
\label{Eq:X_def}
	X = \frac{2 \rho_o g_c R_c}{5} \left( 1-\frac{\rho_o}{\rho_c} \right)  \left[ \left( \alpha + \frac{3}{2} \right) \varepsilon - \beta - \frac{3}{2} - \frac{\rho_c}{\rho_o} \right] S_2 \;.
\end{equation}
From Eq. \eqref{Eq:radial_displacement}, we are now able to write:
\begin{equation}
\label{Eq:epsilon}
	\varepsilon = \frac{
		\frac{19\mu_c}{2\rho_c g_c R_c} + \frac{\rho_o}{\rho_c} \left( 1-\frac{\rho_o}{\rho_c} \right) \left( \beta + \frac{3}{2} \right) 
			+ \left( 1-\frac{\rho_o}{\rho_c} \right)
		}
		{ \left( \alpha + \frac{3}{2} \right) \frac{\rho_o}{\rho_c} \left( 1-\frac{\rho_o}{\rho_c} \right) } \;.
\end{equation}

To treat the anelastic case, we can apply the correspondence principle \citep{Biot..1954}, stipulating that all the developments made so far are still valid if we consider now complex quantities.
Let us then denote by $\tilde{x} $ the Fourier transform of a given quantity $x$.
The complex Love numbers at the core and planet surfaces can be expressed in terms of $\tilde{\varepsilon}$:
\begin{equation}
\label{Eq:k2complex_epsilon_form}
	\tilde{k}_2(R_c) = \frac{3}{2}  \frac{ \tilde{\varepsilon} - 1 + \frac{\rho_c}{\rho_o} }{\alpha \tilde{\varepsilon} - \beta} \;,
	\textrm{ and }\;
	\tilde{k}_2(R_p) = \frac{3}{2}  \frac{ \tilde{\varepsilon} + \frac{2}{3} \, \beta }{\alpha \tilde{\varepsilon} - \beta} \;.
\end{equation}

Thanks to eq. \eqref{Eq:k2complex_epsilon_form}, we are now able to relate the tidal dissipation in the core to the effective tidal dissipation of the planet, given by the imaginary part of the corresponding Love numbers.
Let us denote by $\varepsilon_1$ (resp. $\varepsilon_2$) the real (resp. imaginary) part of $\tilde{\varepsilon}$.
The imaginary parts of the Love numbers $\tilde{k}_2(R_c)$ and $\tilde{k}_2(R_p)$ are
\begin{subequations}
\label{Eq:Im_k2complex_epsilon_form}
\begin{align}
	\Imcro{\tilde{k}_2(R_c)} = - \frac{k_2^c}{Q_c}
		&= \frac{3}{2}  \frac{ \left[ \left( 1 - \frac{\rho_c}{\rho_o} \right) \alpha - \beta \right] \varepsilon_2 }
							 {\left( \alpha \varepsilon_1 - \beta \right)^2 + \left( \alpha \varepsilon_2 \right)^2 } \;, \\
	\Imcro{\tilde{k}_2(R_p)} = - \frac{k_2^p}{Q_p}
		&= - \frac{3}{2}  \frac{ \left( 1 + \frac{2}{3} \alpha \right) \beta \varepsilon_2 }
							 {\left( \alpha \varepsilon_1 - \beta \right)^2 + \left( \alpha \varepsilon_2 \right)^2 } \;.
\end{align} 
\end{subequations}

\begin{subequations}
\label{Eq:Qeff_Gp_formule}
Eq. \eqref{Eq:k2Qc_bicouche} linking these two quantities then becomes
\begin{equation}
\label{Eq:k2Qc_k2Qp_bicouche}
	\frac{k_2^p}{Q_p} = \left( \frac{R_c}{R_p} \right)^5 G_p \frac{k_2^c}{Q_c} \;,
\end{equation}
where $G_p$ takes the form
\begin{equation}
\label{Eq:Gp_expres}
	G_p = \dfrac{\alpha + \dfrac{3}{2}}{\alpha + \dfrac{3}{2} \left( \dfrac{R_c}{R_p} \right)^5}  \;.
\end{equation}
\end{subequations}
First, one may note that this quantity is not related to the factor $F$ that accounts for the enhancement of the tidal deformation of the core due to the presence of the envelope (see Eq.~(28) of \citetalias{RMZL..2012}).
That is in contradiction with \cite{Dermott.etal..1988}, in which, moreover, no derivation of the formula (given here in Eq. \ref{Eq:k2Qc_bicouche}) was detailed.
Second, we remark that $G_p$ does not depend on $\varepsilon$ which links the amplitudes of the deformation of the core and planet surfaces, and thus, neither on the shear modulus of the core $\bar{\mu}_c$.
It only depends on the density contrast between the core and the envelope (through the ratio $\rho_o/\rho_c$) and the core size (through its normalized radius $R_c/R_p$).
This result is coherent with the fact that we have considered a non-dissipative envelope.
Therefore, the factor $G_p$ appears like a quantity that characterizes the transmission of the core's tidal dissipation up to the surface of the planet.
Last, $G_p$ tends to the value $5/2$ of the uniform asymptotic case when either $R_c/R_p \rightarrow 0$ or $\rho_o/\rho_c \rightarrow 1$.

\section{Application to Jupiter- and Saturn-like planets}
\label{Sect:Jup_Sat}

In \citetalias{RMZL..2012}, we used the standard formula (Eq.~\ref{Eq:k2Qc_STD}) to quantify the effective tidal dissipation of Jupiter- and Saturn-like planets when modeled by a two-layer synthetic planet.
In the present paper we show how the use of the refined formula (Eq.~\ref{Eq:Qeff_Gp_formule}) increases its predicted amplitude.

\subsection{Internal structure parameters}
\label{SubSect:Modeles_synthetic}

As explained in \citetalias{RMZL..2012}, we have poor constraints on the internal structure of the giant planets Jupiter and Saturn.
Nonetheless, several models have been developed which give some insight on the plausible density profiles in such planets.
Following \citetalias{RMZL..2012}, we used the models of \cite{Guillot..1999} for Jupiter and \cite{Hubbard.etal..2009} for Saturn, from which we are able to build two-layer synthetic planetary models with a dense central icy/rocky core and a fluid envelope made of hydrogen and helium:
\begin{itemize}
	\item Jupiter-like: $R_c^\mrm{J} = 0.126 \times R_p^\mrm{J}$ and $M_c^\mrm{J} = 6.41 \times M_\mcal{\Phi}$,
	\item Saturn-like: $R_c^\mrm{S} = 0.219 \times R_p^\mrm{S}$ and $M_c^\mrm{S} = 18.65 \times M_\mcal{\Phi}$,
\end{itemize}
where $M_\mcal{\Phi}$ stands for the mass of the Earth.
The most unknown quantities concern the rheology of the predicted core of giant planets.
Even so, one may consider some boundary values defining intervals in which the viscoelastic parameters, of the used Maxwell rheological model, are likely to take their values.
In \citetalias{RMZL..2012}, we estimated such ranges based on our present knowledge of the rheology of the Earth mantle and the icy satellites of Jupiter \citep{Henning.etal..2009,Tobie..2003}.
Recent studies on rocks viscosity at very high pressure \citep{Karato..2011} were also considered.
Thus, in \citetalias{RMZL..2012}, the more realistic values of the viscoelastic parameters were assumed to lie in the ranges: $10^{12} \,\mrm{Pa}\cdot\mrm{s}^{-1} \ie \eta \ie 10^{21} \,\mrm{Pa}\cdot\mrm{s}^{-1}$ for the viscosity, and $4\times 10^9 \,\mrm{Pa} \ie G \ie 10^{11} \,\mrm{Pa}$ for the rigidity, for an unknown mixture of ice and silicates. 
We may also extrapolate the range of values taken by $G$ at very high pressure and temperature by using a simple Steinberg-Cochran-Guinam law \citep{Steinberg.etal..1980}:
\begin{equation}
\label{Eq:loi_SCG_rigidite}
	G(P,T) = G_0 + \frac{\partial G}{\partial P} \frac{P}{(\rho/\rho_0)^{1/3}} + \frac{\partial G}{\partial T} (T-300) \;,
\end{equation}
where $P$, $T$, $\rho$, and $\rho_0$ are respectively the pressure, temperature, density, and density at the reference temperature $T = 300 \,\mrm{K}$.
Assuming the values of $G_0 = 1.66\times 10^{11} \,\mrm{Pa}$, $\partial_P G = 1.56$, and $\partial_T G = -0.020 \,\mrm{GPa}\cdot\mrm{K}^{-1}$ given by \cite{Murakami.etal..2012} for the lower mantle of the Earth, and the rough approximation $(\rho/\rho_0)^{1/3} = 1$, the rigidity of silicates at $T=12300 \,\mrm{K}$ and $P=2000 \,\mrm{GPa}$ (as intermediate values at the upper boundary of the core of Jupiter and Saturn) reaches the value $G = 3\times 10^{12} \,\mrm{Pa}$.
In the following, we will thus assume as the most likely values of the viscoelastic parameters:
\begin{itemize}
	\item for the viscosity: $10^{12} \,\mrm{Pa}\cdot\mrm{s}^{-1} \ie \eta \ie 10^{21} \,\mrm{Pa}\cdot\mrm{s}^{-1}$,
	\item for the rigidity: $4\times 10^9 \,\mrm{Pa} \ie G \ie 10^{12} \,\mrm{Pa}$.
\end{itemize}

\subsection{Impact of the envelope height}
\label{SubSect:Gp_Struct}

\begin{figure}[!htb]
\centering
 \includegraphics[width=\linewidth] {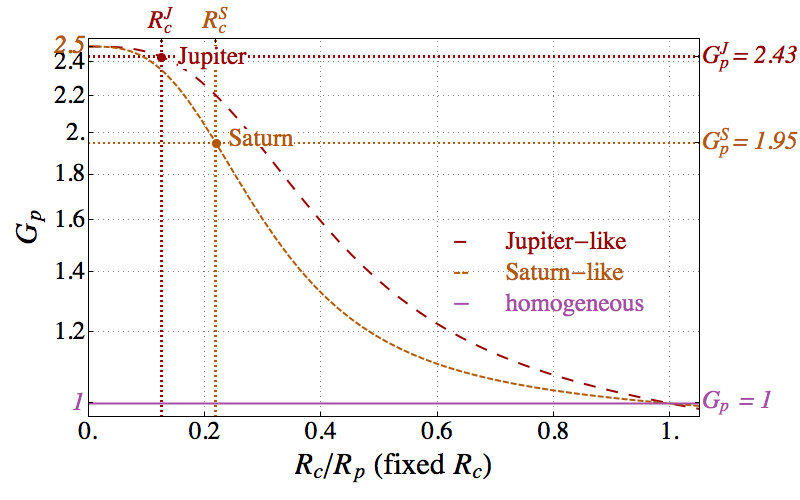}
 \caption{ \label{Fig:Graph_Gp_JS}\it 
 Factor $G_p$ which quantifies the impact of the fluid envelope on the effective tidal dissipation factor $Q_p$, as a function of the envelope height through the ratio $R_c/R_p$.
 }
\end{figure}

Our goal is to quantify the role of the fluid envelope on the effective tidal dissipation.
According to Eq.~\eqref{Eq:Gp_def}, the strength of the associated mechanism is given by the factor $G_p$, represented in Figure~\ref{Fig:Graph_Gp_JS} as a function of the fluid envelope height, for gas giant planets.
 This figure shows that the effective tidal dissipation increases with the height of the envelope, by a factor of $1.95$ to $2.43$ for an envelope at least as high as those of Jupiter and Saturn.

\subsection{Amplitude of the effective tidal dissipation}
\label{SubSect:Qeff_rheology}

\begin{figure}[!htb]
\centering
\begin{minipage}{0.89\linewidth}
	\subfloat[Effective tidal dissipation factor $Q_\mrm{eff}$ of Jupiter, standard formula]
	[Jupiter-like (standard)]
	{\label{Fig:Qeff_rheo_J12} \includegraphics[height=0.498\linewidth]{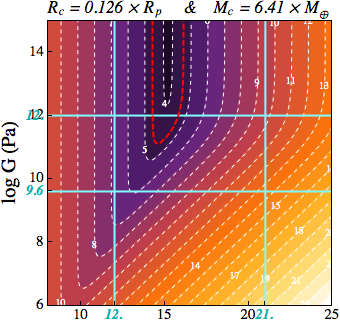}}
	\subfloat[Effective tidal dissipation factor $Q_\mrm{eff}$ of Jupiter, new formula]
	[Jupiter-like (new)]
	{\label{Fig:Qeff_rheo_J14} \includegraphics[height=0.498\linewidth]{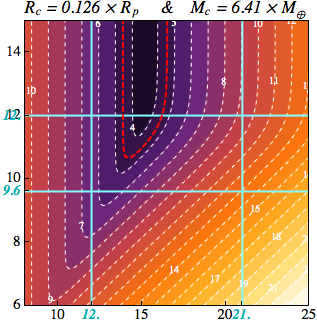}}
	\\
	\subfloat[Effective tidal dissipation factor $Q_\mrm{eff}$ of Saturn, standard formula]
	[Saturn-like (standard)]
	{\label{Fig:Qeff_rheo_S12} \includegraphics[height=0.54\linewidth]{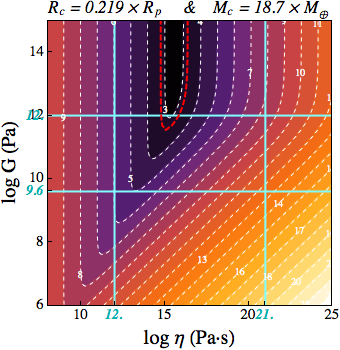}}
	\subfloat[Effective tidal dissipation factor $Q_\mrm{eff}$ of Saturn, new formula]
	[Saturn-like (new)]
	{\label{Fig:Qeff_rheo_S14} \includegraphics[height=0.54\linewidth]{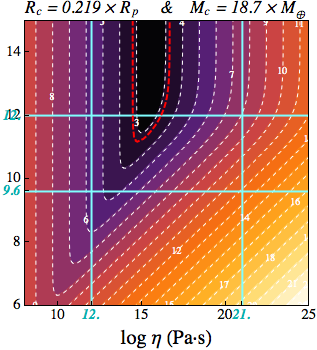}}
\end{minipage}
\hspace{1pt}
\begin{minipage}{0.08\linewidth}
	\centering
	\subfloat{\includegraphics[width=\linewidth]{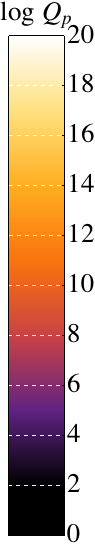}}
\end{minipage}
\caption{\it \label{Fig:Qeff_rheo}
 Effective tidal dissipation factor $Q_p$ of synthetic models of Jupiter (a,c) and Saturn (b,d) as a function of the viscoelastic parameters $G$ and $\eta$: (a,b) use Eq.~\eqref{Eq:k2Qc_STD} and (c,d) use Eq.\eqref{Eq:Qeff_Gp_formule}.
 The red dashed lines give the values of \citet{Lainey.etal..2009,Lainey.etal..2012}.
 The blue lines bound the region of the most  likely rheological values (see text).
  We assume: ${R_p = \{10.97 , 9.14\}} R_{\mcal\Phi}$ ($R_{\mcal\Phi}$ is the Earth radius); ${M_p = \{317.8, 95.16\}} M_{\mcal\Phi}$.
 }
\end{figure}

Using astrometric data, \citet{Lainey.etal..2009,Lainey.etal..2012} determined the tidal dissipation in Jupiter (${Q_\mrm{Jupiter} = (3.56 \pm 0.56) \times 10^4}$) and Saturn (${Q_\mrm{Saturn} = (1.682 \pm 0.540) \times 10^3}$), respectively.
 Fig.~\ref{Fig:Qeff_rheo} shows the amplitude of the effective tidal dissipation factor $Q_p$, as given by Eq.~\eqref{Eq:k2Qc_STD} for Figs~\ref{Fig:Qeff_rheo_J12} and~\ref{Fig:Qeff_rheo_S12}, and Eq.~\eqref{Eq:Qeff_Gp_formule} for Figs~\ref{Fig:Qeff_rheo_J14} and~\ref{Fig:Qeff_rheo_S14}, as a function of the viscoelastic parameters $G$ and $\eta$  of the Maxwell model, for the synthetic models of Jupiter and Saturn built in sec.~\ref{SubSect:Modeles_synthetic}, and compares the results with the observed values.
\captionsetup[subfloat]{position=top,margin=0pt}

 The contour plots of $Q_p$ from Eq.~\eqref{Eq:Qeff_Gp_formule} (Figs.~\ref{Fig:Qeff_rheo_J14} \& \ref{Fig:Qeff_rheo_S14}) are sligthly shifted to the bottom and expanded when compared to those resulting from Eq.~\eqref{Eq:k2Qc_STD} (Figs.~\ref{Fig:Qeff_rheo_J12} \& \ref{Fig:Qeff_rheo_S12}), as in \citet{RMZL..2012}.
Hence, if we only consider the more realistic values of the rheological parameters (as defined in sec.~\ref{SubSect:Modeles_synthetic}), the observed values of tidal dissipation in Jupiter and Saturn are reached for sligthly enlarged ranges of values of $G$ and $\eta$ compared with previous results (Tab.~\ref{Tab:params_rheo_JS}).

\captionsetup[subfloat]{position=top,margin=0pt}
\begin{table}[!h]
\subfloat[Jupiter]
	[Jupiter]
	{
\centering
\begin{tabular}{ | l | l |}
\hline
  Standard model (Eq.~\ref{Eq:k2Qc_STD}) & This paper (Eq.~\ref{Eq:Qeff_Gp_formule}) \\
\hline
 $\intervalleR{G}{1.28\times 10^{11}}{10^{12}}$ Pa & $\intervalleR{G}{4.46\times 10^{10}}{10^{12}}$ Pa \\
 $\intervalleR{\eta}{0.18}{6.23} \times 10^{15} $ Pa$\cdot$s & $\intervalleR{\eta}{0.07}{15.5}\times 10^{15} $ Pa$\cdot$s \\
 \hline
\end{tabular}
}
\\
\subfloat[Saturn]
	[Saturn]
	{
\centering
\begin{tabular}{ | l | l |}
\hline
  Standard model (Eq.~\ref{Eq:k2Qc_STD}) & This paper (Eq.~\ref{Eq:Qeff_Gp_formule}) \\
\hline
 $\intervalleR{G}{3.52\times 10^{11}}{10^{12}}$ Pa & $\intervalleR{G}{1.49\times 10^{11}}{10^{12}}$ Pa \\
 $\intervalleR{\eta}{0.59}{6.58} \times 10^{15} $ Pa$\cdot$s & $\intervalleR{\eta}{0.28}{13.7}\times 10^{15} $ Pa$\cdot$s \\
 \hline
\end{tabular}
}
\caption{\it \label{Tab:params_rheo_JS}Ranges of values of the rheological parameters $G$ and $\eta$ for which the effective tidal dissipation factor $Q_p$ of the two-layer model reaches the observed values.}
\end{table}

\section{Conclusions}

In this work, we revisit the role of the fluid envelope of a two-layer giant planet where the dissipation is assumed to originate from the anelastic core.
We derive an improved formulation of the effective tidal dissipation factor (Eq.~\ref{Eq:Qeff_Gp_formule}) that we express in terms of the tidal dissipation factor of the core obtained by \citetalias{RMZL..2012}.
It improves the modeling of the role of the anelastic core of giant planets as an important source of tidal dissipation.
 When applied to the gas giants Jupiter and Saturn, the amplitude of their effective tidal dissipation is enhanced by about a factor of two, thus keeping the same order of magnitude (see figs.~\ref{Fig:Graph_Gp_JS} and \ref{Fig:Qeff_rheo}).
However, the model takes now into account all the physical mechanisms that act on the tidal dissipation of the core of giant planets.
Finally, this new model has been used recently to compare the relative strength of the different tidal mechanisms that take place in gas giant planets interior, namely the anelastic tidal dissipation in the core \citep{RMZL..2012} and the turbulent friction acting on tidal inertial waves in the convective envelope \citep{Ogilvie.Lin..2004}, showing that the former could slightly dominate the latter \citep{Guenel.etal..2014}.

\begin{acknowledgements}

This work was supported in part by the Programme National de Plan\'etologie (CNRS/INSU), the EMERGENCE-UPMC project EME0911, the CNRS programme {\it Physique the\'eorique et ses interfaces}, the Campus Spatial Universit\'e Paris-Diderot, and the L'Oreal France-UNESCO foundation.

\end{acknowledgements}


\bibliographystyle{aa}
\bibliography{RMZL14}

\end{document}